\documentclass[12pt]{article}
\usepackage{epsfig}
\usepackage{amssymb}
\usepackage{amsfonts}
\def\eV{{\rm e\kern-0.12em V}} \def\GeV{{\rm G}\eV} \def\MeV{{\rm M}\eV}
\def\MSbar{\relax\ifmmode\overline{\rm MS}\else{$\overline{\rm MS}${ }}\fi}
\def\msbar{\relax\ifmmode\overline{\rm MS}\else{$\overline{\rm MS}${ }}\fi}
\def\alphan{\relax\ifmmode{\alpha_{\rm an}}\else{$\alpha_{\rm an}${ }}\fi}
\def\asmz{\relax\ifmmode\bar \alpha_s(M_Z^2)\else{$\bar \alpha_s(M_Z^2)${ }}\fi}
\def\albars{\relax\ifmmode{\bar{\alpha}_s}\else{$\bar{\alpha}_s${ }}\fi}
\def\tildal{\relax\ifmmode{\tilde{\alpha}}\else{$\tilde{\alpha}${ }}\fi}
\def\tildals{\relax\ifmmode{\tilde{\alpha}(s)}\else{$\tilde{\alpha}(s)${ }}\fi}
\def\asQ{\relax\ifmmode\bar{\alpha}_s(Q^2)\else{$\bar{\alpha}_s(Q^2)${ }}\fi}
\newcommand{\beq}{\begin{equation}} \newcommand{\eeq}{\end{equation}}
\begin{document}
\thispagestyle{empty}
\begin{flushright}
JINR preprint E2-2000-46,\\
\medskip

{\it Dedicated to the memory of Vladimir Lengyel}
\end{flushright}
\begin{center}
{\large\sf Toward the correlated analysis of perturbative QCD
observables}         \\
D.V. Shirkov \\
{\it Bogoliubov Laboratory, JINR, 141980 Dubna, Russia\\
e-address: shirkovd@thsun1.jinr.ru}
\end{center} \smallskip
\centerline{\bf Abstract}
{\small We establish direct connection between ghost--free formulations of
RG--invariant
perturbation theory in the both Euclidean and Minkowskian regions. \par
    By combining the trick of resummation of the $\pi^2$--terms for the
invariant QCD coupling and observables in the time-like region with fresh
results on the ``analyticized" coupling $\alpha_{\rm an}(Q^2)$ and
observables in the space-like domain we formulate a self--consistent scheme,
free of ghost troubles. The basic point of this joint construction is the
``dipole spectral relation" (\ref{dipole}) emerging from axioms of local
QFT. \par

 Then we consider the issue of the heavy quark thresholds and devise a
global  scheme for the data analysis in the whole accessible space-like
and time-like domain with various numbers of active quarks.
Observables in both the regions are presented in a form of
non-power perturbation series with improved convergence properties.\par

 Preliminary estimates indicate that this global scheme produces results a
bit different -- on a few per cent level for $\bar{\alpha}_s$ --
from the usual one, thus influencing the total picture of the QCD parameters
correlation.}
\vskip -6mm

{\small  \tableofcontents }
\newpage \addtocounter{page}{-1}
\section{Introduction \label{s1}}
\subsection{Preamble}
 The item of the low energy behavior of a strong interaction attracts more
and more interest along with the further experimental data accumulation.
In the perturbative quantum chromodynamics (pQCD) this behavior is spoiled
by unphysical singularities lying at the three flavour region $(f=3)\,$ and
associated with the scale parameter $\,\Lambda_{f=3} \simeq 350\,$ MeV. In
the ``small energy" and ``small momentum transfer" regions $(\,\sqrt{s},\,
Q\equiv\sqrt{Q^2} \lesssim 3\Lambda\,)$ these singularities
 complicate theoretical interpretation of data. On the other hand, their
existence contradicts some general statements of the local QFT.

 Meanwhile, this issue has a rather elegant solution. As it has been
shown~\cite{rapid96,prl97} (see, also fresh review \cite{ss99tmp}), by
combining three elements: \par
1. Usual Feynman perturbation theory for effective coupling(s) and
observables,\par
2. Renormalizability, i.e., renormalization--group (RG) invariance, and \par
3. General principles of local QFT --- like causality, unitarity, Poincar\'e
invariance and spectrality --- in the form of spectral representations of
K\"allen--Lehmann and Jost--Lehmann--Dyson type \par

it turns out to be possible to formulate an {\it Invariant Analytic Approach}
(IAA) for the pQCD invariant coupling and observables in which the central
theoretical object is a {\it spectral density}. Being calculated by usual,
RG--improved, perturbation theory it defines and relates $\,Q^2$--analytic,
RG-invariant expressions in the both Euclidean and Minkowskian channels. \par
\smallskip

  The IAA obeys several remarkable properties: \par
--- It enables one to obtain modified perturbation expressions for
observables, free of unphysical singularities, poles and cuts,
with behavior correlated in both space-like and time-like
domains.\par
 --- In particular, the IAA results in modified ghost-free expressions for
invariant QCD coupling  $\,\alpha_{\rm an}(Q^2;f)\,$ and $\,\tildal(s;f)\,$
which obey reduced higher--loops and renormalization--scheme
sensitivity~\cite{prl97} -- \cite{nicos}. See, Fig.1.  \par
--- Then, it yields changing the structure of perturbation expansion for
observables:  instead of common power series, as a result of its
integral transformation, there appear non-power asymptotic
series~\cite{lmp99tmp} {\it \`a la Erd\'elyi\/} over the sets of
specific functions ${\cal A}_k(Q^2;f)$ and ${\mathfrak A}_k
(s;f)$. These functions are defined via integral transformations
of related powers $\alpha_s^k(Q^2;f)\,$ in terms of relevant
spectral densities. At small and moderate argument values, they
diminish with the $\,k\,$ growth much quicker than the
corresponding powers $\,\alpha_{\rm an}^k(Q^2;f)\,$ and
$\tildal^k(s;f)\,$, (and even oscillate in the region $ \sqrt{s},
Q\simeq \Lambda $) thus improving essentially the convergence of
perturbation expansion for observables. \par
  We review all these IAA features, important for our further developments,
in the second part of Section \ref{s1}.  \smallskip

  The first purpose of this work is to elucidate relation between the
Radyushkin--Krasnikov--Pivovarov ``pipization" trick\cite{rad,kras} and
the Solovtsov--Milton\cite{ms97} construction of effective $s$--channel
QCD coupling within the IAA scheme. \par
 In the course of this analysis --- see Section \ref{s2} --- we reveal a
spectacular ``distorting mirror"
correlation between analyticized and pipizated invariant QCD coupling in
space-like $\,\alpha_{\rm an}(Q^2;f)\,$ and time-like $\,\tildal(s;f)\,$
regions as well as between corresponding expansion functions
${\cal A}_k(Q^2;f)$ and ${\mathfrak A}_k (s;f)$. See Fig.2. \par \smallskip

  Then, in Section \ref{ss2.3}, we consider an issue of transition across the
heavy quark thresholds, for constructing a ``global" picture valid in the
whole physical region $M_{\tau}\lesssim \sqrt{s}, Q \lesssim M_Z$. \par

It should be noted, that all precedent papers Refs. \cite{rapid96} --
\cite{kras} dealt only with the massless quarks at fixed flavour number,
$\,f\,$, case. This can be justified, to some extent, for analysis inside
a narrow interval of relevant energy $\sqrt{s}\,$ or momentum transfer $Q\,$
values. Meanwhile, the ultimate goal of all the pQCD business is a
correlation of QCD effective coupling values extracted from different
experiments. \par

 To construct the global invariant analytic couplings, one needs recipe
of relating expressions with different $\,f\,$ values. For this goal, we
use the same guideline as previously, at the ``fixed $\,f\,$ case", that
is we start with
adequately defined ``global" spectral functions --- see expression
(\ref{discont}). Here, an essential point is the matching condition that
relates $\Lambda_f\,$ parameters for different fixed flavor number $f\,$
values. We use standard \MSbar prescription ascending to early 80s. 
This results in a smooth global Euclidean $\alphan (Q)\,$, ${\cal A}_k(Q^2)$
and spline--continuous Minkowskian $\tildals\,$,  ${\mathfrak A}_k(s)\, $
expressions. \par \smallskip

  In the concluding Section, using examples of inclusive $\tau$ decay,
$e^+e^- \to$ hadrons annihilation and sum rules, we shortly comment the
possible implication of our new global scheme on perturbative analysis of
QCD processes. \par
  The main results of this work are summed concisely in the Subsection
\ref{ss4.3}.

\subsection{The $s$--channel: early attempts \label{ss1.2}}
 As it is well known, the notion of invariant (or effective) coupling
originally was introduced in the RG treatment\cite{rg56} of
renormalizable QFT. In the RG formalism, invariant coupling
function $\overline{\alpha}$ was defined as a product of
propagator and vertex amplitudes initially related with a product
of real finite Dyson's renormalization constants. This
construction is valid only in the space-like domain \footnote{
Physically, in QED it can be considered as a Fourier transform of
spatial electron charge distribution first discussed by Dirac
\cite{dirac34}.} and can be directly used for analysis of
corresponding observables. However, the RG formalism does not
provide us with analogous object in time-like region. \par

 It is worth noting that sporadic attempts to define the effective
coupling $\alpha(s)$ in the Minkowskian, time-like, domain were
made in late 70s. Omitting an early simple--minded trick with
``mirror reflection" of singular function
$$\alpha_s(Q^2;f)\to\alpha(s;f)\equiv|\alpha_s(-s;f)|,$$ we
mention here the practically simultaneous results of Radyushkin
\cite{rad} and Krasnikov and Pivovarov \cite{kras}. In both the
papers, the integral transformation $ \tildal(s; f)= {\cal\bf R}
[\albars(Q^2; f)]$ reverse  $\,{\bf R}={\cal\bf D}^{-1}$ to
``dipole representation" for the Adler function
\begin{equation}\label{dipole}
D(Q^2) = \frac{Q^2}{\pi}\int^{\infty}_0 \frac{d
s}{(s+Q^2)^2}\,R(s)\,\equiv {\cal\bf D} \left\{ R(s)\right\} \eeq
in terms of an observable $\,R(s)\,$ in the time-like region, has
been used. \par
 In \cite{rad,kras}, as a starting point for observables in the
Euclidean, i.e., space-like domain $\,Q^2 >0$, the perturbation
series
\begin{equation} \label{Dpt}
D_{\rm pt}(Q^2)=1+ \sum_{k\geq 1}^{} d_k\,\albars^k(Q^2; f) \end{equation}
has been assumed. It contains powers of usual, RG summed, invariant
coupling $\albars(Q^2; f)\,$ that obeys unphysical singularities in the
infrared (IR) region around $Q^2 \simeq \Lambda^2_3\,$. \par
    By using the reverse transformation
\begin{equation}\label{contour}
R(s)=\frac{i}{2\pi}\,\int^{s+i\varepsilon}_{s-i\varepsilon}\frac{d
z}{z}\, D_{\rm pt}(-z)\equiv{\bf R}\left[D_{\rm pt}(Q^2)\right]
\end{equation}
these authors arrived at the ``${\bf R}$--transformed" expansion that,
in our notation, reads
\begin{equation}\label{Rpi}
R_{\pi}(s)=1+\sum_{k\geq 1}^{}d_k{\mathfrak{A}}_k(s;f)\,;\quad
~\mathfrak{A}_k(s;f)={\bf R}\left[\albars^k(Q^2;f)\right]
\,.\end{equation}

 For example
$${\bf R}\left[\frac{1}{l}\right]=\frac{1}{2}-\frac{1}{\pi}
\arctan\frac{L}{\pi}\,;\quad\mbox{with}\quad l=\ln\frac{Q^2}{\Lambda^2}\,
;\quad L=\ln\frac{s}{\Lambda^2}\,,$$
$${\bf R}\left[\frac{\ln l}{l^2}\right]=
\frac{\ln\left[\sqrt{L^2+\pi^2}\right]+1-L{\bf R}\left[1/l\right]}{L^2+
\pi^2}\,;\quad {\bf R}\left[\frac{1}{l^2}\right]=\frac{1}{L^2+\pi^2}\,.$$
 This yields\footnote{This expression we give in the form equivalent to
that one used in \cite{ms97}. In papers \cite{rad,pivo} it was given in
 an another form, non-adequate at $L\leq 0$. See, also \cite{necholom}.}
\beq \label{tildal1}
\tildal^{(1)}(s; f)=\frac{1}{\beta_0}\left[\frac{1}{2}-\frac{1}{\pi}
\arctan\frac{L}{\pi}\right]\:; \quad\beta_0=\frac{33-2f}{12\pi} \,.\eeq
 At the two--loop iterative case with
$$ \beta_{[f]}\albars^{(2)}(Q^2; f)=\frac{1}{l}-b_f\frac{\ln
l}{l^2} \,,\quad \beta_{[f]}\equiv\beta_0
\,;~\quad\beta_1=\frac{102-38f}{12\pi}\:, \quad b_f
=\frac{\beta_1}{\beta_0^2}\,\:,$$
 by combining ${\bf R}\left[1/l\right]-b_f{\bf R}[\ln l/l^2]$ one obtains
explicit expression for the ``iterative" two-loop effective $s$--channel
coupling $\tildal^{(2)}(s; f)=\mathfrak{A}_1^{(2)}(s;f)\,,$
$$
\tildal^{(2)}_{iter}(s;f)=\left(1-\frac{b_f L}{L^2+\pi^2}
\right)\tildal^{(1)}(s;f) +\frac{b_f}{\beta_{[f]}}\frac{\ln\left[
\sqrt{L^2+\pi^2}\right]+1}{L^2+\pi^2}\,. $$

Obtained $\tildal^{(1)}$ and $\tildal^{(2)}_{\rm iter}$ are monotonous
functions with finite IR limit, free of $\Lambda$--singularity which is
``screened" by resummed ``$\pi^2$--terms". Non-singular expressions for
higher functions ${\mathfrak A}_k$ could be constructed in the same way.
\smallskip

  The positive feature of this construction was an automatic summation
of the so--called ``$\pi^2$ -- terms" that ``screen" unphysical
singularities and observed\cite{rad} property $$ \left({\bf
R}\left[\albars^{k+1}\right]\right)^{1/(k+1)} < \left({\bf R}
\left[\albars^{k}\right]\right)^{1/k}\/ $$ that improves the
convergence of perturbation series. \par
   However, there was one essential drawback. The dipole transformation
(\ref{dipole}), that is supposed to be reverse to ${\bf R}\,$, being
applied to (\ref{Rpi}) does not return us to the input (\ref{Dpt})
$$
{\cal\bf D}\left\{R_{\pi}(s)\right\}={\cal\bf D}\left\{{\bf R}
\left[D_{\rm pt}\right]\right\} \neq D_{\rm pt}(Q^2)\,~~~\Rightarrow ~~~~
 {\cal\bf D}\cdot {\bf R} \neq {\mathbf I}\,, $$
as far as the unphysical singularities of $\albars(Q^2;f)\,$ and of its
powers are incompatible with analytic properties in the complex $\,Q^2\,$
plane of the integral in the r.h.s. of (\ref{dipole}).

 Resolution of this issue came 15 years later with the IAA. The
``missing link" is the analyticization transformation.

\subsection{Analyticization in the $Q^2$--channel \label{ss1.3}}
Operation 
$$F(Q^2)\,\to F_{\rm an}(Q^2)\,= {\cal\bf A}\cdot F(Q^2)\,$$ has
been introduced~\cite{rapid96} in terms of the K\"all\'en--Lehmann
representation and correlates with analytic properties of the
Adler function contained in eq.(\ref{dipole}).\par

 Generally, this transformation is defined for a function $F$ that should
be analytic in the $Q^2$ plane with a cut along the negative part of the
real axis. In our case, this function could be either invariant coupling
$\albars$ itself \footnote{As it has been explained in detail in the first
papers~\cite{rapid96,prl97} on the IAA, the QCD invariant coupling,
according to general properties of local QFT, should satisfy the
K\"all\'en--Lehmann spectral representation. For the original analysis of
this issue see Ref. \cite{ilya65}.} or its power, or some series in its
powers. \par \smallskip

 Operation ${\cal\bf A}\,$ consists of two elements:\\
 -- use the K\"allen--Lehmann representation $$F_{\rm
an}(Q^2)=\frac{1}{\pi}\int\limits_{0}^{\infty}\frac{d
\sigma}{\sigma + Q^2}\, \rho_{\rm pt}(\sigma)\quad \,\mbox{with}$$
-- the spectral density defined via straightforward continuation
of $F$ on the cut $$\rho_{\rm pt}(\sigma)=\Im\:F(-\sigma) \,.$$
\smallskip

 A couple of comments are in order. 
 \begin{itemize}
\item Operation ${\cal\bf A}$, being applied to the usual
coupling\footnote{For the time being, we consider the massless
case with a fixed number $\,f\,$ of effective quark flavors in the
\msbar scheme. For the transition between the regions with
different $\,f\,$ values, see Section \ref{ss2.3}.}
$\,F=\albars(Q^2;f)\,$, results in the analyticized coupling
\begin{equation} \label{spectr}
\alpha_{\rm an}(Q^2; f)=\frac{1}{\pi}
\int\limits_{0}^{\infty}\frac{d \sigma} {\sigma + Q^2}\,
\rho(\sigma; f) \,\,;~~ \rho(\sigma; f)=\Im \albars(-\sigma;
f)\,\eeq
which is a smooth monotonic function free of unphysical singularities,
with a finite value at the origin
$$
\,\alpha_{\rm an}(0; f)= 1/\beta_{[3]}\simeq 1.4\,$$
which is remarkably independent (see, e.g., \cite{prl97}) of higher
loop contributions.

 Here, $\rho$ is defined as an imaginary part of the usual, RG invariant,
effective coupling $\albars$ continued on the physical cut.
\item Operation ${\bf A}$, applied to power perturbation series (\ref{Dpt})
for an observable $D_{\rm pt}(Q^2)$, produces a non-power series
\begin{equation}\label{Dan}                                        
D_{\rm an}(Q^2;f)=1+\sum_{k\geq 1} d_k\,{\cal A}_k(Q^2;f)\:;  \quad
\alpha_{\rm an}(Q^2;f)= {\cal A}_1 (Q^2;f)
\end{equation}
with
\begin{equation}\label{defAk}
{\cal A}_k(x;f)=\frac{1}{\pi}\int\limits_{0}^{\infty}\frac{d\sigma}
{\sigma+x}\:\rho_k(\sigma;f)\,;~~\rho_k(\sigma;f)
=\Im\left[\albars^k(-\sigma;f) \right]\,. \end{equation}
 \end{itemize}

  For example,
$$ {\bf A} \cdot \left[\frac{1}{l}\right] = \frac{1}{l} +
\frac{1}{1-e^l}\,,~~~~~ {\bf A} \cdot \left[\frac{1}{l^2}\right] =
\frac{1}{l^2} + \frac{e^l}{(1-e^l)^2} \,,~ \dots$$%
 that is
\begin{equation}\label{an1Q}                 
 {\bf A}\cdot\alpha_s^{[1]}(Q^2; f)=\alpha_{\rm an}^{[1]}(Q^2;f)
= \frac{1}{\beta_0}\left[\frac{1}{\ln(Q^2/\Lambda^2}
+\frac{\Lambda^2}{\Lambda^2-Q^2}\right]\end{equation}
and so on.

  Here, in the invariant analytic coupling \alphan, the
$\Lambda$--pole is compensated by power term containing the
non-perturbative
$Q^2/\Lambda^2=(Q^2/\mu^2)\exp(1/\beta_0\alpha_{\mu})$ structure.
\par
 Properties of the invariant analytic functions ${\cal A}_k$, free of
ghost troubles, and non-power expansion (\ref{Dan}) have been discussed
in papers \cite{lmp99tmp}. They are quite similar to those for
$\mathfrak{A}_k$ and expansion (\ref{Rpi}) --- see below.

\subsection{Summary of the IAA \label{ss1.4}}
 Here, we repeat in brief basic definitions of the Invariant Analytic
Approach.\par
 First, one has to transform the usual singular invariant coupling
$$ \albars(Q^2;f)\to {\cal\bf A}\cdot \albars(Q^2;f) =
\alpha_{\rm an}(Q^2;f)\, $$
into the analyticized one, free of ghost singularities in the space-like
region. \par
 Second, with the help of the operation ${\bf R}$, one defines\cite{ms97}
invariant coupling $\tildal(s;f)$ in the time-like domain
\beq\label{9}
\alphan(Q^2;f)\to\tildal(s;f)={\bf R}\left[\alphan\right] =
\int\limits^{\infty}_{s}\frac{d\,\sigma}{\sigma}\rho(\sigma; f)\,\eeq
with spectral density $\rho\,$ defined in (\ref{spectr}).
\smallskip

 Here, we have a possibility of reconstructing the Euclidean,
$Q^2$--channel, invariant coupling $\alphan(Q^2; f)$ from the Minkowskian,
$s$--channel, one $\tildal(s; f)\,$ by the dipole transformation
\begin{equation}\label{al-dipole}
\alphan(Q^2; f) =\frac{Q^2}{\pi}\int^{\infty}_0\frac{d s}
{(s+Q^2)^2}\, \tildal(s; f)\,\equiv {\cal\bf D}\left\{\tildal(s;
f)\right\}\;.\eeq

  For instance, substituting $\tildal^{(1)}(s; f)$ into the integrand,
one obtains after integration by parts

$$\frac{Q^2}{\pi\beta_0}\int^{\infty}_{0}\frac{d\sigma}{(\sigma+
Q^2)^2}\cdot\left(\frac{1}{2}-\frac{1}{\pi}\arctan
\frac{\ln(\sigma/\Lambda^2)}{\pi}\right)=$$ $$
\frac{Q^2}{\pi\beta_0}\int^{\infty}_{0}\frac{d\sigma}{(\sigma+Q^2)}
\,\,\frac{1}{\ln^2(\sigma/\Lambda^2)+\pi^2}=\alpha^{(1)}_{\rm
an}(Q^2, f)$$ precisely in the form (\ref{an1Q}).
 This simple calculation elucidates the connection between the ghost--free
expressions both in the $s$-- and $Q^2$--channels. They are connected also
by the reverse transformation
 $ \tildal^{(1)}(s;f) ={\bf R}\left[\alphan^{(1)}(Q^2;f)\right]\,. $

On the Fig. 1 we give a concise summary of the IAA results for
invariant analytic couplings $\alphan(Q^2,3)$ and $\tildal(s,3)$
calculated for one-- , two-- and three--loop cases
in both the Euclidean and Minkowskian domains. \par

 \begin{figure}\label{fig1}
 \unitlength=1mm
 \begin{picture}(0,100)
 \put(10,1){%
 \epsfig{file=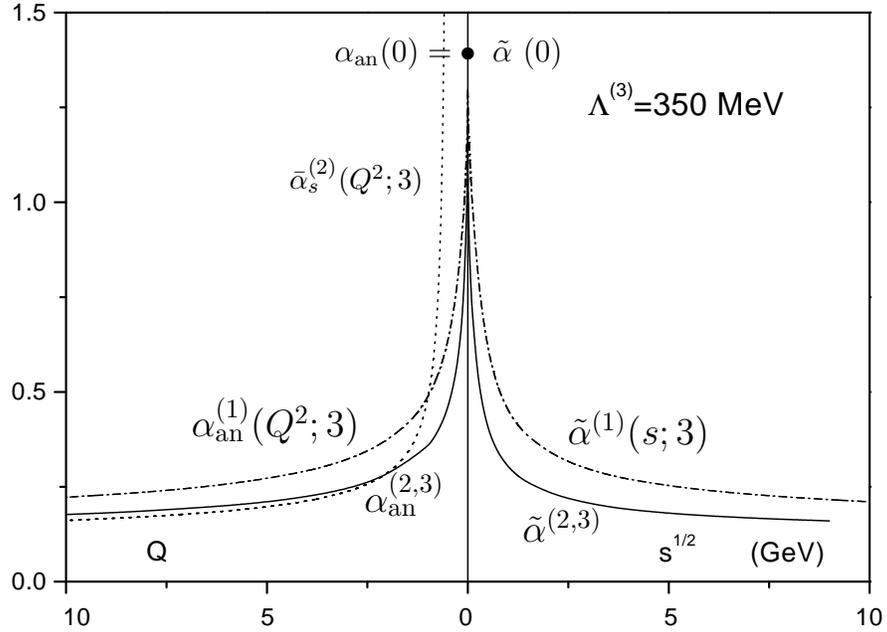,width=14cm}%
 }
\put(82.7,87){$\bullet$}%
\put(66,87){$\alphan(0)=$}%
\put(87,87){\tildal(0)}%
 \put(60,70){\small $\bar\alpha_s^{(2)}(Q^2;3)$}%
 \put(47,37){\large $\alpha_{\rm{an}}^{(1)}(Q^2;3)$}%
 \put(70,27){\large $\alpha_{\rm an}^{(2,3)}$}%
 \put(97,36){\large $\tilde{\alpha}^{(1)}(s;3)$}%
 \put(91,23){\large $\tildal^{(2,3)}$}%
\end{picture}
 \caption{\sl Space-like and time-like invariant analytic couplings in
 a few GeV domain} %
 \end{figure}
 Here, dash--dotted curves represent one-loop IAA approximations
(\ref{tildal1}) and (\ref{an1Q}). Solid IAA curves are based on exact
two-loop solutions of RG equations\footnote{
 As it has been recently established, the exact solution to two-loop RG
differential equation for the invariant coupling can be expressed
in terms of a special function $W$, the Lambert function, defined
by relation $ W(z) e^{W(z)}=z \,,$ with an infinite number of
branches $\,W_n(z)$. For some detail of analyticized 
solutions expressed in terms of Lambert function --- see Refs.
\cite{ggk98,badri98,badri99,Lambert}.} and approximate three--loop
solutions in the \msbar scheme. Their remarkable coincidence
(within the 1--2 per cent limit) demonstrates reduced sensitivity
of the IAA with respect to the higher--loops effects in the whole
Euclidean and Minkowskian regions from IR till UV limits.
\par
 For comparison, by dotted line we also give usual \asQ two-loop
effective QCD coupling with a pole at $Q^2=\Lambda^2\,.$ \par

 As it has been shown in \cite{prl97,ss99tmp,MSS-PL97-tau}, relations
parallel to eqs.(\ref{9}) and (\ref{al-dipole}) are valid for
powers of the pQCD invariant coupling. This can be resumed in the
form of a self-consistent scheme.
\section{Self-consistent scheme for observables\label{s2}}
\subsection{Relations between Euclidean and Minkowskian \label{ss2.1}}
 First, one has to transform usual power perturbation series (\ref{Dpt})
of the $\,Q^2\,$ domain
$$ {\bf I.}~~~~~D_{\rm pt}(Q^2)\to D_{\rm an}(Q^2)=
{\cal\bf A}\cdot D_{\rm pt}(Q^2)\, $$
 into the non-power one (\ref{Dan}). \par \smallskip

Second, with the help of the operation ${\bf R}$, one introduces
$$ {\bf II.}~~~~~
 D_{\rm an}(Q^2)\to R_{\pi}(s)={\bf R}\left[D_{\rm an}(Q^2)\right]$$
the s--channel non-power expansion $R_{\pi}(s)$ (\ref{Rpi}) with
\begin{equation}\label{m-s}                                          
\mathfrak{A}_k(s)=\int\limits^{\infty}_{s}\frac{d \sigma}{\sigma}
\rho_k(\sigma)\,\,; ~~\rho_k(\sigma)
=\Im\left[\alpha_s^k(-\sigma)\right]\,. \eeq
\smallskip

 The third element is the closure of the scheme that is provided by the
operation (\ref{al-dipole})
$$ {\bf III.}~~~~~~~
R_{\pi}(s)\to D_{\rm an}(Q^2)={\cal\bf D}\left\{R_{\pi}(s)\right\}\, $$
reverse to {\bf II.} \smallskip

  In the other words, to enjoy self-consistency
$\,{\bf R}\cdot{\cal\bf D} = {\cal\bf D}\cdot {\cal\bf R}={\bf 1}\,, $ one
should {\it abandon completely} the usual effective coupling $\alpha_s(Q^2)$
and power series $D_{\rm pt}\,$, eq.(\ref{Dpt}), applying operations
${\cal\bf R}$ and ${\cal\bf D}= {\cal\bf R}^{-1}$ only to IAA invariant
couplings \alphan, \tildal and to non-power expansions $D_{\rm an}$
and $R_{\pi}\,$.
\subsection{Expansion of observables over non-power sets
$\left\{{\cal A}\right\}$ and $\left\{{\mathfrak A}\right\}$\label{ss2.2}}
 To realize the effect of transition from expansion over the ``traditional"
power set $$\,\left\{\albars^k(Q^2,f)\right\}
\,=\,\asQ,\,\albars^2,\, \dots \albars^k \dots \, $$ to expansions
over non--power sets in the space-like and time-like domains $$
\left\{{\cal A}_k(Q^2,f)\right\}\,=\,\alpha_{\rm an}(Q^2,f),\, {\cal
A}_2(Q^2,f) ,\,{\cal A}_3 \dots~;~\left\{{\mathfrak A}_k(s,f) \right\}
=\,\tilde{\alpha}(s,f),\,{\mathfrak A}_2(s,f),\,{\mathfrak A}_3\dots
\,,$$ it is instructive to learn properties of the latters.

  In a sense, both non-power sets are similar: \par
--- They consist of functions that are free of unphysical singularities. \par
--- First functions, the new effective couplings, ${\cal A}_1 =
\alpha_{\rm an}$ and ${\;\mathfrak A}_1 = \tilde{\alpha}\/$ are
monotonically decreasing. In the IR limit, they are finite and equal
$\alpha_{\rm an}(0,3)=\tilde{\alpha}(0,3)\simeq 1.4\,$ with the same
infinite derivatives. Both have the same leading term
$\,\sim 1/\ln x\/$ in the UV limit.\par

--- All other functions (``effective coupling powers") of both the sets
start from the zero IR values ${\cal A}_{k\geq 2}(0,f)=
{\mathfrak A}_{k\geq 2}(0,f)=0$ and obey the UV behavior
$\sim 1/(\ln x)^k$\, corresponding to $\albars^k(x)$. They are no longer
monotonous. The second functions $\,{\cal A}_2\,$ and $\,{\mathfrak A}_2\,$
are positive with maximum around $\;s, Q^2 \sim \Lambda^2$. Higher functions
$\,{\cal A}_{k\geq 3}\,$ and $\,{\mathfrak A}_{k\geq 3}\,$ oscillate in the
region of low argument values and obey $\,\/k-2\/\,\;$ zeroes. \par

  Remarkably enough, the mechanism of liberation of unphysical
singularities is quite different. While in the space-like domain
it involves non-perturbative, power in $Q^2$, structures, in the
time-like region it is based only upon resummation of the
``$\pi^2$ terms". Figuratively, (non-perturbative !) {\it
analyticization} in the $Q^2$--channel can be treated as a
quantitatively distorted reflection (under $Q^2\to s=- Q^2$) of
(perturbative) {\it``pipization"} in the $s$--channel. This effect
of ``distorting mirror" first discussed in \cite{mo98} is
illustrated on figures 1 and 2. \smallskip

  Summarize the main results essential for data analysis. Instead of
power perturbative series in the space-like $$ \hspace{30mm} D_{\rm
pt}(Q^2)=1+d_{\rm pt}(Q^2)\:;~~ d_{\rm pt}(Q^2)= \sum_{k\geq
1}^{}d_k\,\albars^k(Q^2;f) \hspace{29mm} (\ref{Dpt}a) $$ and
time-like regions $$ R_{\rm pt}(s)=1+r_{\rm pt}(s)\,;\quad r_{\rm
pt}(s)=\sum_{k\geq 1}r_k\, \tildal^k(s;f)\,;\quad
(r_{1,2}=d_{1,2}, r_3=d_3- d_1\frac{\pi^2\beta_{[f]}^2}{3}),$$
 one has to use asymptotic expansions (\ref{Dan}) and (\ref{Rpi})
$$d_{\rm an}(Q^2)= \sum_{k\geq 1}d_k\, {\cal A}_k(Q^2,f) \,;~\quad
r_{\pi}(s)=\sum_{k\geq 1}d_k\, {\mathfrak A}_k(s,f) $$ with the same
coefficients $d_k$ over non-power sets of functions $\left\{{\cal
A}\right\}$ and $\left\{{\mathfrak A}\right\}$.

\subsection{Global formulation  \label{ss2.3}}
 To apply the new scheme for analysis of QCD processes, one has to formulate
it ``globally", in the whole experimental domain, i.e., for regions with
different values of a number $f$ of active quarks. For this goal, we revise
the issue of the threshold crossing. \par \smallskip

{\sf Threshold matching \label{sss2.3.1}} In a real calculation, the
procedure of the threshold matching is in use. One of the simplest is the
matching condition in the massless \msbar scheme\cite{match}
\begin{equation}\label{Q2match}
\albars(Q^2=M^2_f; f-1) = \albars(Q^2=M^2_f; f)\eeq
related to the mass squared $M_f^2$ of the f-th quark. \par
   This condition allows one to define a ``global" function
$\,\albars(Q^2)\,$ consisting of the smooth parts
$$
\albars(Q^2)=\,\albars(Q^2;f)\quad\mbox{at}\quad M^2_{f-1}\leq
Q^2\leq  M^2_f $$
s and continuous in the whole space-like interval of positive
$\,Q^2\,$ values with discontinuity of derivatives at the matching
points. We call such a functions as the {\it spline--continuous}~
ones.

  At the first sight, any massless matching, yielding the spline--type
function, violates the analyticity in the $Q^2\,$ variable, thus
disturbing the relation between the $s$-- and
$Q^2\,$--channels\footnote{Any massless scheme is an approximation
that can be controlled by the related mass--dependent
scheme~\cite{dv81}. Using such a scheme, one can
devise~\cite{mikh} a smooth transition across the heavy quark
threshold. Nevertheless, from the practical point of view, it is
sufficient (besides the case of data lying in close vicinity of
the threshold) to use the spline--type matching (\ref{Q2match})
and forget about the smooth threshold crossing.}. \par

 However, in the IAA, the original power perturbation series (\ref{Dpt}) with
its unphysical singularities and possible threshold non-analyticity has no
direct relation to data, being a sort of a ``raw material" for defining
spectral density. Meanwhile, the discontinuous density is not dangerous.
Indeed, expression of the form
\begin{equation}\label{discont}                                          
\rho_k(\sigma)=\rho_k(\sigma; 3) + \sum_{f\geq 4}^{}\theta(\sigma-M_f^2)
\left\{\rho_k(\sigma; f)-\rho_k(\sigma; f-1) \right\} \end{equation}
with $~\rho_k(\sigma; f)= \Im\,\albars^k(-\sigma, f)$
 defines, according to (\ref{defAk}) and (\ref{m-s}), the smooth global
\begin{equation} \label{globalAQ}
{\cal A}_k(Q^2) =\frac{1}{\pi}\int\limits_{0}^{\infty} \frac{d\sigma}
{\sigma+x} \:\rho_k(\sigma)
\end{equation}
 and spline--continuous global
\begin{equation} \label{globalAs}
\mathfrak{A}_k(s)=\int\limits^{\infty}_{s}\frac{d \sigma}{\sigma}
\rho_k(\sigma)\, \end{equation}
 functions \footnote{Here, by eqs.(\ref{globalAQ}),(\ref{globalAs}) and
(\ref{discont}) we introduced new ``global" effective invariant couplings
and higher expansion functions different from the previous ones with fixed
$f$ value.}.\par

 We see that in this construction the role of the input perturbative invariant
coupling $\albars(Q^2)$ is twofold. It provides us not only with spectral
density (\ref{discont}) but with matching conditions (\ref{Q2match})
relating $\,\Lambda_f\,$ with $\,\Lambda_{f+1}\,$ as well.  \par
  Note that the matching condition (\ref{Q2match}) is tightly related
\cite{match,mikh} to the renormalization procedure. Just for this profound
reason we keep it untouched (compare with Ref.~\cite{mo98}). \smallskip

{\sf The $s$-channel: shift constants\label{sss2.3.2}} As a
practical result, we now observe that the ``global" $s$--channel coupling
$\,\tildal(s)\,$ and other functions ${\mathfrak A}_k(s)$,
generally, differs of effective coupling with fixed flavor number
$f$ value $\,\tildal(s;f)\,$ and ${\mathfrak A}_k(s;f)$ by a
constants. For example, at $M^2_5\leq s\leq M^2_6$ $$\tildal(s)\,=
\,\int\limits^{\infty}_{s}\frac{d \sigma}{\sigma}\rho(\sigma)\,=
\,\int\limits^{M^2_6}_{s}\frac{d \sigma}{\sigma}\rho(\sigma;5)\,+
\,\int\limits^{\infty}_{M^2_6}\frac{d
\sigma}{\sigma}\rho(\sigma;6)\,= \tildal(s; 5)\,+c(5)\,. $$
Generally,
\begin{equation}\label{alpha-s-f}
\tildal(s)\,=\tildal(s; f)\,+c(f)\,\quad \quad \mbox{at}\quad
\quad M^2_f\leq s \leq M^2_{f+1}\,  \eeq
with {\it shift constants} $c(f)$ that can be calculated in terms of
integrals over $\rho(\sigma; f+n) \,\,\, n\geq 1 \,$ with additional reservation
$\,c(6)=0\,$ related to the asymptotic freedom condition. \par More specifically,
$$
c(f-1)=\tildal(M^2_f; f)-\tildal(M^2_f; f-1)+c(f)\,\:,\quad c(6)=0 \:. $$
  These $\,c(f)\,$ reflect the $\,\tildal(s)\,$ continuity at the
matching points $\,M^2_f$.  \par

  Analogous shift constants
\begin{equation}
{\mathfrak A}_k(s)\,=\,{\mathfrak A}_k(s; f)\,+{\mathfrak c}_k(f) \,\quad
\mbox{at} \quad M^2_f\leq s \leq M^2_{f+1}\,\end{equation}                  
are responsible for continuity of higher expansion
functions. Meanwhile, $\,{\mathfrak c}_2(f)\,$ relates to
discontinuities of the ``main" spectral function (\ref{discont}).

 The one-loop estimate with $\,\beta_{[f]}\rho(\sigma;f) =
\left\{\ln^2(\sigma/\Lambda^2_f)+ \pi^2\right\}^{-1}\,$,
\beq\label{sc}
 c(f-1)-c(f) =\frac{1}{\pi\beta_{[f]}}\arctan
\frac{\pi}{\ln \frac{M^2_f} {\Lambda_f^2}}
-\frac{1}{\pi\beta_{[f-1]}}\arctan\frac{\pi} {\ln \frac{M^2_f}
{\Lambda_{f-1}^2}} \simeq \frac{17-f}{54}\albars^3(M_f^2) ~\eeq
and traditional values of the scale parameter $\,\Lambda_{3},
\Lambda_4 \sim 350-250$ \MeV \ reveals that these constants $$
c(5)\simeq 3.10^{-4}\:;~c(4)\simeq 3.10^{-3}\:;~c(3)\sim
0.01\,,\:{\mathfrak c}_2(f)\simeq 3\,\alpha(M_f^2)\,c\,(f) \,$$
are essential at a few per cent level for $\,\tildal\,$ and at ca
10\% level for the $\,{\mathfrak A}_2\,$.  \par

 This means that the quantitative analysis of some $s$--channel events
like, e.g., $e^+e^-$ annihilation~\cite{ss99tmp}, $\tau$--lepton
decay~\cite{MSS-PL97-tau} and charmonium width~\cite{kras} at the $f=3\,$
region should be influenced by these constants. \par
\smallskip

{\sf Global Euclidean functions \label{sss2.3.3} }
  On the other hand, in the Euclidean, instead of the spline-type function
$\albars\,$, we have now continuous, analytic in the
whole $\,Q^2>0\,$ domain, invariant coupling defined, along with
(\ref{globalAQ}), via the spectral integral
\begin{equation} \label{an-spectr}
\alpha_{\rm an}(Q^2)=\frac{1}{\pi} \int\limits_{0}^{\infty}\frac{d \sigma}
{\sigma + Q^2}\; \rho(\sigma) \end{equation}
with the discontinuous density $\rho(\sigma)$ (\ref{discont}). \par

  Unhappily, here, unlike for the time-like region, there is no possibility
of enjoying any more explicit expression for $\,\alpha_{\rm an}(Q^2)\,$
even in the one-loop case. Moreover, the Euclidean functions
$\,\alpha_{\rm an}$ and $\,{\cal A}_k\,$, being considered in a
particular $f$--flavour region $\,M^2_f\leq Q^2 \leq M^2_{f+1}\,$, do depend
on all $\,\Lambda_3\,, \dots ,\,\Lambda_6\,$ values simultaneously. \par

 Nevertheless, the real difference from the $f=3$ case, numerically, is not
big at small $Q^2\,$ and in the ``few GeV region", for practical
reasons, it could be of importance . \medskip

          \begin{figure}[]\label{fig2}
 \unitlength=1mm
  \begin{picture}(0,100)
  \put(10,1){%
  \epsfig{file=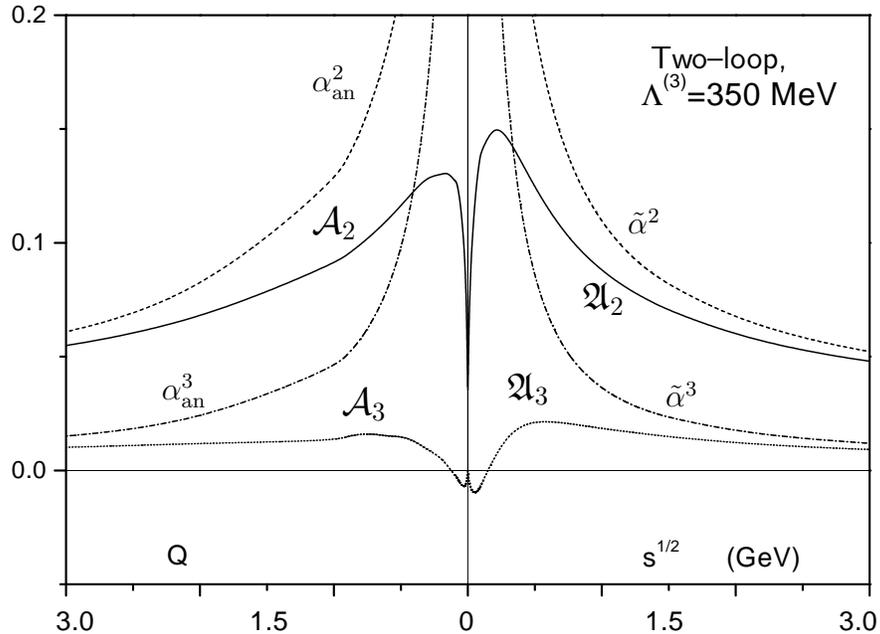,width=14.0cm}%
  }
\put(108,86){\sf Two--loop,}%
  \put(63,83){$\alpha_{\rm an}^2$}%
  \put(63,64){\large\bf ${\cal A}_2$}%
  \put(43,42){$\alpha_{\rm an}^3$}%
  \put(67,40){\bf\large ${\cal A}_3$}%
  \put(105,64){$\tilde{\alpha}^2$}%
  \put(99,54){\large ${\mathfrak A}_2$}%
  \put(110,41){$\tilde{\alpha}^3$}%
  \put(89,42){\large ${\mathfrak A}_3$}%
  \end{picture}
 \caption{\sl  ``Distorted mirror symmetry" for global
expansion functions}
 \end{figure}

 This situation is illustrated by Fig. 2. Here, by thick solid curves
 with maxima around $\sqrt{s}, Q \equiv \Lambda\,$, we draw expansion functions
${\cal A}_2$ and ${\mathfrak A}_2\,$ in a few GeV region. Thin solid lines zeroes
around $\Lambda$ and negative values below, represent ${\cal A}_3$ and
${\mathfrak A}_3\,$.  For comparison, we give also second and third powers
of relevant analytic couplings \alphan and \tildal.

All these functions correspond to exact two--loop solutions
expressed in terms of Lambert function \footnote{Details of these
calculations will be published elsewhere. The assistance of D.S.
Kurashev and B.A. Magradze in calculation curves
 with  Lambert functions is gratefully acknowledged.}.

\section{Correlation of experiments \label{s4}}
  Another quantitative effect stems from the non-power structure of
the IAA perturbative expansion. It is also emphasized at the few GeV region.

\subsection{The $s$--channel \label{ss4.1} }
 To illustrate the qualitative difference between our global scheme and
other practice of data analysis, we first consider the $f=3$ region.  \medskip

  {\sf Inclusive $\tau$ decay.} The IAA scheme with fixed $f=3$ was
used in Ref.~\cite{MSS-PL97-tau} for analysis of the inclusive
$\tau$--decay. Here, the observed quantity, the $\tau$ lepton time of
half--decay, depends on the integral of the $s$--channel matrix element
over the region $0< s < M_{\tau}^2$. As a result of the 2--loop IAA
analysis of the experimental input $R_{\tau}=3.633$~\cite{caso}, the value
$\tildal^{(2)}(M_{\tau}^2)= 0.378\,$ has been obtained that has to be
compared with related result of usual analysis $\albars^{(3)}(M_{\tau}^2)
=0.337\,$. This shift $\Delta \alpha\simeq 0.04$ resulted in a rather big
change in the extracted $\Lambda$ value. Meanwhile, some part of this shift
can be ``absorbed" by the shift constant $\,c(3)\,.$
\smallskip

 The process of {\sf Inclusive $e^+e^-$ hadron annihilation} provides us with
an important piece of information on the QCD parameters.
In the usual treatment, (see, e.g., Refs.\cite{caso, bardin}) the basic
relation looks like
\begin{equation}\label{Rtrad}
\frac{R(s)}{R_0} =1+ r(s)\,; \quad r(s)=\frac{\albars(s)}{\pi}+r_2\,
\albars^2(s)+ r_3\,\albars^3(s)\,. \end{equation}
 Here, the numerical coefficients $\, r_1=1/\pi
=0.318\,,~\,r_2=0.142\,,~r_3=- 0.413\,$ (related to the $f=5$ case) are not
diminishing. However, a rather big negative $r_3$ value comes mainly from
the $\,-r_1\pi^2\beta^2_{[5]}/3$ contribution equal to $-0.456$. Instead of
(\ref{Rtrad}), with due account of (\ref{Rpi}), we now have
\begin{equation}\label{r-new}
r(s)= 1+\frac{\tildal(s)}{\pi}+d_2\,{\mathfrak
A}_2(s)+d_3\:{\mathfrak A}_3(s)\:;\end{equation} with reasonably
decreasing coefficients $\,d_1=0.318\,;~d_2=0.142 \,
;~d_3=0.043\,,$ the mentioned $\pi^2$ term of $r_3$ being
``swallowed" by $\,\tildal(s)\,$\footnote{This term contributes about
$8.10^{-4}$ into the $r(M^2_Z)$  and, correspondingly, 0.0025 into
the extracted $\albars(M^2_Z)$ value. This means, that the main part of
the ``traditional three-loop term" $r_3 \albars^3$ in the r.h.s. of
(\ref{Rtrad}) being of the one--loop origin is essential for the
modern quantitative analysis of the data. In particular, it should be
taken into the account even in the so-called NLLA which is a
common approximation for the analysis of events at $\sqrt{s}=M_Z\,.$
 } .\par
 Now, the main difference between (\ref{r-new}) and (\ref{Rtrad})
is due to the term $\,d_2\,{\mathfrak A}_2\,$ standing in the place of
$~d_2\,\tildal^2$. The difference can be estimated by adding into
(\ref{Rtrad}) the structure $\,r_4\,\alpha^4 \,$ with $\,r_4
=d_2\,\beta_{[5]}^2\pi^2 \simeq -0.62.$ This effect could be essential in
the region of $\tildal(s) \simeq 0.20-0.25$.

\subsection{The $Q^2$--channel \label{ss4.2}}
{\sf The $Q^2$--channel: Bjorken and GLS sum rules.}
In the paper \cite{mss98bj}, the IAA has been applied to the Bjorken sum
rules. Here, one has to deal with the $Q^2$--channel at small transfer
momentum squared $\,Q^2 \lesssim 10\,\GeV^2\,$.
 Due to some controversy of experimental data, we give here only a part of
the results of \cite{mss98bj}.
     For instance, using data of the SMC Collaboration \cite{smc97} for
$Q_0^2=10\,\GeV^2$ the authors obtained $\alpha_{\rm an}^{(3)}(Q_{0}^{2})=
0.301\,$ instead of $\alpha_{\rm pt}^{(3)}(Q_{0}^{2})=0.275\,$. \smallskip

 In the Euclidean channel, instead of power expansion like (\ref{Dpt}),
we typically have
 \begin{equation}
d(Q^2)=\frac{\alpha_{\rm an}(Q^2)}{\pi} +d_2\,{\cal A}_2(Q^2)+
 d_3\,{\cal A}_3(Q^2)\:. \end{equation}
 Here, the modification is related to non-perturbative power
structures behaving like $\,\Lambda^2/Q^2\,$ at $\,Q^2 \gg\Lambda^2\,$.
 As it has been estimated above, these corrections could be essential
in a few \GeV \ region.

  \par The same remark could be made with respect to analysis of the
Gross--Llywellin-Smith sum rules of \cite{mss99gls}.
\smallskip

 {\sf Some comments} are in order:

---  We see that, generally, the extracted values of $\alpha_{\rm
an}$ and of $\tilde{\alpha}\,$ are both slightly greater
in a few GeV region than the relevant values of $\albars$
for the same experimental input. This corresponds to the
above-mentioned non-power character of new asymptotic expansions
with a suppressed higher-loop contribution. \par

---  At the same time, for equal values of
$\,\alphan(x_*)=\tildal(x_*) =\albars(x_*)\,$, the analytic scale
parameter $\Lambda_{\rm an}\,$ values extracted from $\,\alphan\,$
and $\,\tildal\,$ are a bit greater than that $\,\Lambda_{\msbar}\,$ taken
from $\albars$. This feature is related to a ``smoother" behavior of
both the regular functions $\alphan$ and $\tildal\,$ as compared
to the singular $\albars$. \par \smallskip

\subsection{Conclusion \label{ss4.3}}
 To summarize, we repeat once more our main points.\smallskip

 1. We have formulated the 
self-consistent scheme for analyzing data both in the space-like and
time-like regions.
\smallskip

 The fundamental equation connecting these regions is the dipole
spectral relation (\ref{dipole}) between renormalization--group
invariant non-power expansions $D_{\rm an}(Q^2)$ and $R_{\pi}(s)$.
\par

  Just this equation, equivalent to the K\"allen--Lehmann
representation, is responsible for non-perturbative terms in the
$\/Q^2\,$--channel involved into $\alphan(Q^2)$ and non-power
expansion functions $\left\{{\cal A}_k(Q^2)\right\}$. These terms,
non-analytic in the coupling constant $\alpha$, are a counterpart
to the perfectly perturbative $\,\pi^2$--terms effectively summed
in the $s$--channel expressions \tildals and $\left\{{\mathfrak
A}_k(s)\right\}$.\smallskip

2. As a by-product, we ascertain a new qualitative feature of the
IAA, relating to its non-perturbativity in the $Q^2$--domain. It
can be considered as a {\sf minimal nonperturbativity} or minimal
non-analyticity\footnote{Compatible with the RG invariance and the
$Q^2$ analyticity --- compare with \cite{dv76}.} in $\alpha$ as
far as it corresponds to perturbativity in the $s$--channel. \par

 Physically, it implies that minimal non-perturbativity cannot be
referred to any mechanism producing effect in the
$s$--channel.\smallskip

 3. The next result relates to the correlation between regions with
different values of the effective flavor number $f\,$. Dealing
with the massless \msbar renormalization scheme, we argue that the
usual perturbative QCD expansion provides our scheme only  with
step--discontinuous spectral density (\ref{discont}) depending
simultaneously on different scale parameters $\,\Lambda_f\,;~f=
3,\dots,6\,$ connected by usual matching relations.\par

  This step--discontinuous spectral density yields, on the one hand, 
smooth analytic coupling $\alpha_{\rm an}(Q^2)\,$ and higher functions
$\left\{{\cal A}_k(Q^2)\right\}\,$ in the space-like region
 --- eq.(\ref{globalAQ}).\par
 On the other hand, it produces the  spline--continuous invariant
coupling $\,\tilde{\alpha}(s)\,$  and functions $\left\{{\mathfrak
A}_k(s)\right\}$ in the time-like region
--- eq.(\ref{globalAs}).

  As a result, the global expansion functions $\left\{{\cal A}_k(Q^2)
\right\}\,$ and $\left\{{\mathfrak A}_k(s)\right\}\,$ differ both from
the that ones $\left\{{\cal A}_k(Q^2;f)\right\}\,$ and
$\left\{{\mathfrak A}_k(s;f)\right\}\,$ with a fixed value of a
flavour number.\smallskip

4. Thus, our global IAA scheme  uses common invariant coupling
$\,\albars(Q^2, f)\,$ and matching relations, only as
an input. Practical calculation for an observable now involves expansions
over the sets $\,\left\{{\cal A}_k(Q^2)\right\}\,$ and $\,\left\{{\mathfrak
A}_k(s)\right\} \,$, that is non-power series with usual numerical
coefficients $\,d_k\,$ obtained by calculation of the relevant Feynman
diagrams. \par
  This means that, generally, one should 
check the accuracy of the bulk of extractions of the QCD parameters 
from diverse  ``low energy" experimental data. Our preliminary estimate
shows that such a revision could influence the rate of their correlation.
\smallskip

5. Last but not least. As it has been mentioned in our recent publications
~\cite{prl97,ss99tmp}, the IAA obeys an immunity with respect to
higher loop and renormalization scheme effects. \par
  Now, we got an additional insight into this item related to observables
and can state that 
the perturbation series for an observable in the IAA have better
convergence properties (than in usual RG--summed perturbation theory)  in
both the $s$-- and $Q^2\,$-- channels.
\medskip

{\bf\large Acknowledgements}
\smallskip

 The author is indebted to D.Yu.~Bardin, N.V.~Krasnikov, B.A.~Magradze,
S.V. Mi\-k\-hai\-lov, A.V. Radyushkin, I.L. Solovtsov and O.P. Solovtsova
for useful discussion and comments. This work was partially supported by
grants of the Russian Foundation for Basic Research (RFBR projects Nos
99-01-00091 and 00-15-96691), by INTAS grant No 96-0842 and by INTAS-CERN
grant No 2000-377. \vspace{-2mm}

\addcontentsline{toc}{section}{References}

\end{document}